# Exploring the Jarzynski Equality
# for the Harmonic Oscillator

by

Ronald F. Fox
Regents' Professor Emeritus
Georgia Institute of Technology
Atlanta, Georgia
May 10, 2022

## Abstract

Twenty-five years ago [1], Christopher Jarzynski published a paper in which he asserted: the work done, $W$, in driving a system from state $A$ to state $B$, characterized by the Helmholtz free energies $F^A$ and $F^B$, satisfies the *equality*:

$$< \exp[-\beta W] > \; = \exp[-\beta(F^B - F^A)]$$

in which $< ... >$ denotes an average over an ensemble of realizations of the stochastic $W$ along the path from A to B. The *equality* is significant and unexpected. So is the statement that the equality is *independent of the **rate** of change* from state $A$ to state $B$.

Almost ten years ago, I presented two papers in which the contraction of the description from full phase space to coordinate space was made. This was motivated by the large difference in time scales for momenta relaxation and coordinate relaxation at temperatures near 300 K and processes at the nanoscale in space and time. The Jarzynski equality (JE) can be reconsidered in this contracted picture.

The example to be considered here is a macromolecular harmonic oscillator with a Hooke's spring constant, $k$, that during the time interval $(0, \tau)$ linearly changes from $k_1$ to $k_2$. When the transition is performed *sufficiently slowly*, we obtain JE in the form:

$$< \exp[-\beta W] > \; = \sqrt{k_1/k_2}$$

However, when the transition is performed more rapidly the equality is lost.

# Introduction

The choice of the harmonic oscillator as an example for the study of a claim in physics is attractive because everything that can be known is believed to be known already. This is so for the classical case, the quantum case, and the thermodynamic case. In this paper, the harmonic oscillator is chosen to investigate the eponymous thermodynamic equality (JE) discovered by Christopher Jarzynski in 1997. For this purpose, we will ask how JE is justified when we change the spring constant, $k$, linearly in time. Oddly enough, the oscillator turns out to be an exceptional case. While a version of JE is justified, we can show that it is invalidated for sufficiently fast rates of change.

From its inception in 1997, JE has generated a lot of resistance. My purpose here is not to provide a comprehensive account of that debate. I refer the reader to just a few representative papers [2],[3],[4],[5]. The purpose here is to present the details for a particular example that is thought to be representative. While the criticisms have come from many sources Jarzynski has published numerous rebuttals and clarifications. Each time greater clarity has been achieved. Experiments, albeit difficult experiments with limited statistics, have been published supporting the truth of JE. By now, JE is generally accepted as correct and has laid the foundation for a new field of study, the indirect measurement of the Helmholtz free energy for macromolecular states.

The foundation for the approach presented below may be found in two of my papers on thermo-stated systems [6],[7]. Instead of working in full phase space, as have most other researchers in this area, I have contracted out the momenta by integration, yielding a description in coordinate space. This is a consequence of the vast difference in time scales for momenta relaxation (Langevin time scale for the Maxwell distribution) and for coordinate relaxation (thermal conductivity, attenuation of sound, diffusivity time scales for the Boltzmann distribution) at temperatures around 300 K. It is also a result of the high heat capacity and strong thermal conductivity of water, the most abundant molecular species in the systems I consider.

# Markov Approximation to the Fundamental Equation

Using projection operator techniques in concert with boson operator representations of the dynamical operators in the coordinate picture, it is possible to obtain a non-Markovian fundamental equation, without use of any additional approximations. Such an equation is given in Eq. (43) of [7]. The Markov approximation to this equation is given by (in one spatial coordinate for simplicity)

(1)
$$\partial_t R(r,t) = D \partial_r (\partial_r + \beta U') R(r,t)$$

$R(r,t)$ is the probability distribution for the position $r$ at time $t$ with diffusion constant $D$ and force $-U'$ for conservative potential $U$. $\beta$ is the inverse of the Boltzmann constant times the temperature, $k_B T$. This equation can be thought of as the Fokker-Planck equation associated with a particular Langevin equation. It is also referred to as the Smoluchowski equation. Throughout, the temperature is that of the reservoir. First, change $r$ to $x$ so we remember that we are in just one spatial coordinate. The corresponding Langevin equation is

(2)
$$\frac{d}{dt} x = -D\beta U' + g(t)$$

in which $g(t)$ is a Gaussian fluctuating force with zero average and second moment

(3)
$$<g(s)g(s')> = 2D\delta(s - s')$$

which exhibits the white noise limit. This is not **the** Langevin equation for the velocity of a Brownian particle but is instead **a** Langevin equation because of its generic form and represents the standard Brownian motion Langevin equation for the velocity in the diffusion limit (small Reynold's number). The standard Langevin equation is concerned with the relaxation of the velocity of a particle given by the Maxwell distribution. On a timescale long compared to the timescale of **the** Langevin equation the description is one of coordinate diffusion of interacting particles and relaxation to the Boltzmann distribution.

The potential function is chosen to be that for a harmonic oscillator, centered at the origin,

(4)
$$U = \frac{1}{2} k x^2$$

in which $k$ is the Hooke's law spring constant. If this were the whole picture, the equilibrium Helmholtz free energy would be $\beta F = -\ln\left(\sqrt{\frac{2\pi}{\beta k}}\right)$.

To test JE, we introduce a time dependent piece to the potential given by
(5)
$$k \to k_1 + (k_2 - k_1)\frac{t}{\tau}$$

As $t$ goes from 0 to $\tau$, $k$ goes from $k_1$ to $k_2$. $\tau$ serves as the rate determiner (the *switching* time denoted by $t_s$ in [1]). The mutual consistency of Eqs. (1-3) even with a time dependent $k$, follows from N. G. van Kampen's lemma [8].

The definition of the work done, *on* or *by* the system, through contact with a thermal reservoir is crucial to these considerations. The rebuttal to some of the critics has been that they did not precisely use Jarzynski's definition [3]. Jarzynski's definition of the work done is
(6)
$$W(\tau) := \int_0^\tau dt \, \frac{\partial}{\partial t} U[x(t), t]$$

where the partial derivative acts on the free $t$ as in Eq. (5), but not the $t$ inside $x(t)$. $W$ is a *functional* of $x(s)$. The notation is inadequately conveying the proper meaning. The stochastic path, $x(t)$, is labeled by the time $t$ that covers the entire interval $(0, \tau)$. This means that $W(\tau)$ depends on the entire path, $x(t)$, during time interval $(0, \tau)$. In Eq. (4), $U$ directly depends on $x(t)$ only. At time $t$ the potential has the value $U(x)$ for $x(t)$, but in Eq. (6), $W(\tau)$ depends on the entire path and should be written as a functional of $x(t)$ through $U$ using the notation $U[x(*)]$ where the square brackets denote a *functional*, as distinct from a function with round brackets, as in $U(x)$, and the $(*)$ denotes a dependence on time but not a particular time, rather the entire time for the whole path.

Some critics object to Jarzynski's definition because it seemingly converts a simple *function* into a *functional* for no apparent reason. On the other hand, using Jarzynski's definition in Eqs. (4-6), gives
(7)
$$W_J(\tau) := \int_0^\tau ds \, \frac{1}{2\tau}(k_2 - k_1)x^2(s)$$

This is in harmony with the idea that the stochastic work done is not a state function and must depend on the path. I agree with this desire, but I find that it is a natural consequence of the stochastic nature of the formulation and does not require additional imposition. Instead, we will use the definition for the work done during the transition to be given by the most obvious choice, the part of the total potential that causes the transition in $k_1 \to k_2$ as $t \to \tau$.

(8)
$$W_F(t) = \frac{1}{2}(k_2 - k_1)\frac{t}{\tau}x^2(t) \to \frac{1}{2}(k_2 - k_1)x^2(\tau)$$

In this expression, $x(t)$ is a stochastic path and $x(t)$ means the path dependent value of the stochastic $x$ as determined by the evolution from the initial time to time $t$. The stochasticity is generated by $g$ as in Eq. (2). It is the reason we must use a functional for the work term.

Our task is to decide the validity of JE

(9)
$$\langle \exp[-\beta W] \rangle = \sqrt{k_1/k_2}$$

where the averaging is over the stochastic paths $x(s)$. First, we must find the solutions to Eq. (2) for the complete potential, but only for times in $(0, \tau)$. These solutions will be gaussian normal distributions. In Eqs. (7, 8) the square of $x(s)$ appears. The square of a gaussian normal distribution is **not** a gaussian normal distribution. Nevertheless, the average in Eq. (7) can be determined. This is because we know the averages of all powers of a gaussian variable, and the result has a simple combinatorial formula. First, we will consider the Jarzynski definition of the work.

## Solutions to the Langevin Equation

A solution to the Langevin equation

(10)
$$\frac{d}{ds}x = -D\beta\left(k_1 + (k_2 - k_1)\frac{s}{\tau}\right)x + g(s)$$

$$:= -\alpha(s)x + g(s)$$

in the interval $(0, \tau)$ is given by

(11)
$$x(s) = \exp\left[-D\beta\left(k_1 s + (k_2 - k_1)\frac{s^2}{2\tau}\right)\right] x(0)$$

$$+\exp\left[-D\beta\left(k_1 s + (k_2 - k_1)\frac{s^2}{2\tau}\right)\right]\int_0^s dr\, \exp\left[D\beta\left((k_1 r + (k_2 - k_1)\frac{r^2}{2\tau}\right)\right] g(r)$$

$$:= \exp[-A(s)]\, x(0) + \exp[-A(s)] \int_0^s dr\, \exp[A(r)] g(r)$$

explicitly defines $\alpha(s)$ and $A(s)$ and requires $A(s) = \int_0^s dr\, \alpha(r)$. Taking the $s$−derivative of Eq. (11) demonstrates that it is the unique solution to Eq. (10), an *inhomogeneous initial value equation*. This solution is linear in $g$, and linear in $x(0)$, making it gaussian, because $g$ is assumed to be gaussian and because the thermal distribution for the initial position of a harmonic oscillator, $x(0)$, is also gaussian (quadratic potential). One consequence is that the doubly averaged $x(s)$, once with respect to $g(r)$, and once over the initial value of $x(0)$, is zero, as are all odd order averages.

The presence of $g(r)$ requires the explicit presence of all values of $r$ in $(0, s)$ and makes the last line above a functional.

We are now ready to write the left-hand-side of Eq. (9) in terms of Eq. (7) and the solution for $x(s)$ in Eq. (10) is given by (Jarzynski definition)
(12)

$$\langle \exp[-\beta W_J(\tau)]\rangle = \left\langle \sum_{n=0}^{\infty} \left(-\frac{\beta}{2\tau}(k_2 - k_1)\right)^n \frac{1}{n!}\left(\int_0^\tau ds\, x^2(s)\right)^n\right\rangle$$

$$= \sum_{n=0}^{\infty} \left(-\frac{\beta}{2\tau}(k_2 - k_1)\right)^n \frac{1}{n!}\left\langle\left(\int_0^\tau ds\, x^2(s)\right)^n\right\rangle$$

For a gaussian, we know how to average each of its powers. Because the first and second moments of a gaussian determine each of its moments, the average of its powers also reduces to the first two moments, multiplicatively. I quote the well-known formula for the present case (all odd moments vanish)
(13)

$$\left\langle \left(\int_0^\tau ds\, x^2(s)\right)^n \right\rangle = \frac{(2n)!}{n!\, 2^n} \left\langle \left(\int_0^\tau ds\, x^2(s)\right) \right\rangle^n$$

We see that the average of the nth power of the integral of $x^2(s)$ is expressed in terms of the nth power of the average of the integral of $x^2(s)$.

After reaching this place in a prequel of this paper, Jarzynski challenged the validity of Eq. (13) during a long email exchange between us. Its validity requires that the *integral of the square of a gaussian is also the square of a gaussian*. The challenge was successful. I withdrew the paper, and I went back to beginning where I chose Eq. 8) over Eq. (7), since, as explained above, it already contained the path dependent quality of a functional work term. By doing so, the troublesome integral was eliminated. Now we get (Fox definition)

(14)
$$\langle \exp[-\beta W_F(t)] \rangle = \left\langle \sum_{n=0}^\infty \left(-\frac{\beta t}{2\tau}(k_2 - k_1)\right)^n \frac{1}{n!} \left(x^2(t)\right)^n \right\rangle$$

$$= \sum_{n=0}^\infty \left(-\frac{\beta t}{2\tau}(k_2 - k_1)\right)^n \frac{1}{n!} \left\langle \left(x^2(t)\right)^n \right\rangle$$

$$= \sum_{n=0}^\infty \left(-\frac{\beta t}{2\tau}(k_2 - k_1)\right)^n \frac{1}{n!} \frac{(2n)!}{n!\, 2^n} \langle x^2(t) \rangle^n$$

Clearly, the integral over $s$ in Eq. (13) is replaced by the multiplicative factor of $t$ in Eq. (14). No modifications to the Langevin equation have been made, such as in Eq. (6). Before we finish the calculation of the last average, notice that the combinatorial structure created by this derivation is very basic (for all $|C| < \frac{1}{2}$).

(15)
$$\sum_{n=0}^\infty (-C)^n \frac{1}{n!} \frac{(2n)!}{n!\, 2^n} = \frac{1}{\sqrt{1 + 2C}}$$

One begins to see how an averaged exponential can turn into a reciprocal square-root, as in Eq. (9).

# The Last Average

The square of the solution Eq. (11) is (omitting cross-terms with average 0)

(16)
$$x^2(s)$$
$$= exp[-2A(s)]\, x^2(0)$$
$$+ exp[-2A(s)] \int_0^s dr \int_0^s dr'\, exp[A(r)] exp[A(r')] g(r) g(r')$$

Eq. (3) makes $r = r'$ and generates a factor of $2D$. The equilibrium average for $x^2(0)$ is $1/\beta k_1$. Therefore, for the averaged square (two averages) we get (the cross-correlations vanish so that nothing arises from products of $x(0)$ and $g$ if either factor occurs an odd number of times.)

(17)
$$\{\langle x^2(s) \rangle\} = exp[-2A(s)] \left( \frac{1}{\beta k_1} + \int_0^s dr\, exp[2A(r)] 2D \right)$$

The integral above is doable

(18)
$$\int_0^s dr\, exp[2A(r)] 2D$$

$$= \frac{exp\left[\frac{D\beta\tau k_1^2}{(k_1 - k_2)}\right] \sqrt{\pi\tau} \left( Erf\left[\frac{\sqrt{D\beta\tau}\, k_1}{\sqrt{(k_1 - k_2)}}\right] - Erf\left[\frac{\sqrt{D\beta}(k_2 s + k_1(\tau - s))}{\sqrt{\tau}\sqrt{(k_1 - k_2)}}\right] \right)}{2\sqrt{D\beta}\sqrt{(k_1 - k_2)}} 2D$$

$$= \frac{exp\left[-\frac{D\beta\tau k_1^2}{(k_2 - k_1)}\right] \sqrt{\pi D\beta\tau} \left( Erf\left[\frac{\sqrt{D\beta\tau}\, k_1}{\sqrt{(k_1 - k_2)}}\right] - Erf\left[\frac{\sqrt{D\beta\tau}\left(k_2 \frac{s}{\tau} + k_1\left(1 - \frac{s}{\tau}\right)\right)}{\sqrt{(k_1 - k_2)}}\right] \right)}{\beta\sqrt{(k_1 - k_2)}}$$

Putting Eqs. (17, 18) together gives

(19)
$$\{\langle x^2(s)\rangle\} = \exp[-2A(s)]\left(\frac{1}{\beta k_1} + \int_0^s dr\, \exp[2A(r)]2D\right)$$

$$= \exp\left[-2D\beta\tau\left(k_1\frac{s}{\tau} + (k_2-k_1)\frac{s^2}{2\tau^2}\right)\right]\left(\frac{1}{\beta k_1}\right.$$

$$\left. + \frac{\exp\left[-\frac{D\beta\tau k_1^2}{(k_2-k_1)}\right]\sqrt{\pi D\beta\tau}\left(Erf\left[\frac{\sqrt{D\beta\tau}k_1}{\sqrt{(k_1-k_2)}}\right] - Erf\left[\frac{\sqrt{D\beta\tau}\left(k_2\frac{s}{\tau} + k_1\left(1-\frac{s}{\tau}\right)\right)}{\sqrt{(k_1-k_2)}}\right]\right)}{\beta\sqrt{(k_1-k_2)}}\right)$$

Of particular interest is the result for $s = \tau$. (If we were implementing Jarzynski's definition here, then an numerical integral over $s$ would be requireded and does not exist explicitly as a known function.)

(20)
$$\{\langle x^2(\tau)\rangle\} = \exp[-D\beta\tau(k_1+k_2)] \times$$

$$\left(\frac{1}{\beta k_1}\right.$$

$$\left. + \frac{\exp\left[-\frac{D\beta\tau k_1^2}{(k_2-k_1)}\right]\sqrt{\pi D\beta\tau}\left(Erf\left[\frac{\sqrt{D\beta\tau}k_1}{\sqrt{(k_1-k_2)}}\right] - Erf\left[\frac{\sqrt{D\beta\tau}k_2}{\sqrt{(k_1-k_2)}}\right]\right)}{\beta\sqrt{(k_1-k_2)}}\right)$$

To better understand these expressions, we will select the values for the quantities in Eq. (20) representative of the heads of kinesin molecules, a variety of motor protein with a long elastic tether. I choose the temperature to be 290 Kelvin so that $\beta = 1/k_B T = .25 \times 10^{14}\, ergs^{-1}$. The Langevin relaxation time for a kinesin head is $\tau_R = 1.14 \times 10^{-12}\, seconds$. The mass is $8 \times 10^{-20}\, grams$. The Stokes drag formula for a radius of 3.5 nm (to a physicist a kinesin head is approximately a sphere [9]) is $\alpha = 7 \times 10^{-8} \frac{gm}{s}$. These values imply a diffusion constant of $D = \frac{kT}{\alpha} = \frac{4}{7} \times 10^{-6} \frac{cm^2}{s}$. The Hooke constant for the kinesin neck linker, that is elastic, is estimated to be $k = 6.9 \frac{gm}{s^2}$. The time, $\tau$, is the duration of the transition. I am choosing $\tau \in (10^{-10}s, 1s)$. Notice that in Eq. (20) the parameters $D, \beta$ and $\tau$ are almost always together as $D\beta\tau$. For the values of $D$ and $\beta$ selected above, and for the choice $\tau = 7 \times 10^{-8}s$, the product is $D\beta\tau = 1$, in $cgs$ units. We suspect that any reasonable experiment would take a time longer, even much longer, than this.

We want to put Eq. (20) into Eq. (14). Notice that the $\beta's$ in the denominators of Eq. (20) cancel out the $\beta's$ in the numerators of Eq. (14). This produces
(21)
$$\{\langle \exp[-\beta W_F(\tau)]\rangle\} = \left\{\left\langle \sum_{n=0}^{\infty} \left(-\frac{\beta}{2}(k_2 - k_1)\right)^n \frac{1}{n!} (x^2(\tau))^n \right\rangle\right\}$$

$$= \sum_{n=0}^{\infty} \left(-\frac{1}{2}(k_2 - k_1)\right)^n \frac{1}{n!} \frac{(2n)!}{n! \, 2^n} (X^2)^n$$

wherein $X^2$ is defined by
(22)
$$X^2 := \exp[-D\beta\tau(k_1 + k_2)] \times \left(\frac{1}{k_1}\right.$$

$$\left.+ \frac{\exp\left[-\frac{D\beta\tau k_1^2}{(k_2-k_1)}\right]\sqrt{\pi D\beta\tau}\left(Erf\left[\frac{\sqrt{D\beta\tau}k_1}{\sqrt{(k_1-k_2)}}\right] - Erf\left[\frac{\sqrt{D\beta\tau}k_2}{\sqrt{(k_1-k_2)}}\right]\right)}{\sqrt{(k_1-k_2)}}\right)$$

Several comments are in order. In Eq. (14) only a single type of averaging is implied whereas in Eq. (21) the double averaging is implied. Since $x(t)$ is the sum of two independent averages, one over $g$ and one over $x(0)$, the double averages of all even powers of their sums reduces to powers of the sum of their second moments. This is proved in appendix A. The arguments of the error functions (Erf) are purely imaginary because we have implicitly assumed the $k_2 > k_1$ and that makes the factor $\sqrt{(k_1 - k_2)}$ imaginary. For purely real arguments, the values of the error functions are bounded by asymptotes, $\pm 1$, but for imaginary arguments, their imaginary values have unbounded moduli. The error functions are divided by the square root of a negative number since the integrations that produced Eq. (22) assumed that $k_2 > k_1$. Thus, the expressions in Eq. (22) are real, albeit with large moduli in some cases.

Eqs. (21-22) suggest the following structure for computing the values of $W_F(\tau)$.
(23)
$$\{\langle \exp[-\beta W_F(\tau)]\rangle\} = \sum_{n=0}^{\infty} \left(-\frac{1}{2}(k_2 - k_1)\right)^n \frac{1}{n!} \frac{(2n)!}{n! \, 2^n} (X^2)^n$$

The quantities $\beta, D, k_1, k_2$ are fixed. Only $\tau$ is variable. Recall that for the quantities in our example, $\beta D \tau = 1 \frac{s^2}{gm}$ for $\tau = 7 \times 10^{-8} s$. The combination, $\beta D \tau$ accounts for all instances of $\beta, D, \tau$, leaving only the $k$'s. In the example, $k_1 = 6.9 \frac{gm}{s^2}$, and for specificity let us choose $k_2 = 8.1 \frac{gm}{s^2}$. The remaining combinations of $k'$s in Eq. (22) yield
(24)
$$k_1 + k_2 = 15.0 \frac{gm}{s^2} \qquad \frac{1}{k_1} = 0.145 \frac{s^2}{gm} \qquad \frac{k_1^2}{(k_2 - k_1)} = 5.878 \frac{gm}{s^2}$$

$$\frac{k_1}{\sqrt{(k_1 - k_2)}} = 6.299i \sqrt{\frac{gm}{s^2}} \qquad \frac{k_2}{\sqrt{(k_1 - k_2)}} = 7.394i \sqrt{\frac{gm}{s^2}}$$

$$\sqrt{(k_1 - k_2)} = 1.095i \sqrt{\frac{gm}{s^2}}$$

The values for the Erf functions for $\tau = 7 \times 10^{-8} s$ are
(25)
$$Erf\left[\frac{\sqrt{D\beta\tau}k_1}{\sqrt{(k_1 - k_2)}}\right] - Erf\left[\frac{\sqrt{D\beta\tau}k_2}{\sqrt{(k_1 - k_2)}}\right]$$

$$= Erf[6.299i] - Erf[7.394i]$$

$$= -4.266 \times 10^{22} \, i$$

This number has a very large modulus. The value of $\tau$ used corresponds to a very fast time for the transition. A slower transition is created by increasing the size of $\tau$, say to $10^{-3}\ s$. This makes the argument of the Erf functions one hundred times bigger. The asymptotic expansion for the Erf function of an imaginary argument, $x \to \infty\ i$, is

(26)
$$Erf[x] \sim 1 + \frac{-1}{x\sqrt{\pi}} Exp[-x^2]$$

Therefore, we get

(27)
$$X^2 \sim \exp[-D\beta\tau(k_1 + k_2)] \times \left(\frac{1}{k_1}\right.$$

$$\left. + \frac{\exp\left[-\frac{D\beta\tau k_1^2}{(k_2 - k_1)}\right]\sqrt{\pi D\beta\tau}\left(\frac{-1}{\sqrt{\pi}}\left(\frac{Exp\left[-\frac{D\beta\tau}{k_1 - k_2}k_1^2\right]}{\frac{\sqrt{D\beta\tau}k_1}{\sqrt{(k_1 - k_2)}}} - \frac{Exp\left[-\frac{D\beta\tau}{k_1 - k_2}k_2^2\right]}{\frac{\sqrt{D\beta\tau}k_2}{\sqrt{(k_1 - k_2)}}}\right)\right)}{\sqrt{(k_1 - k_2)}}\right)$$

$$= \exp[-D\beta\tau(k_1 + k_2)]\frac{1}{k_1} - \exp[-D\beta\tau(k_1 + k_2)]\frac{1}{k_1} + \frac{1}{k_2}$$

$$= \frac{1}{k_2}$$

Putting this into Eq. (23) gives

(28)

$$\sum_{n=0}^{\infty} \left(-\frac{1}{2}(k_2 - k_1)\right)^n \frac{1}{n!} \frac{(2n)!}{n! \, 2^n} \left(\frac{1}{k_2}\right)^n$$

$$= \frac{1}{\sqrt{1 + \frac{k_2 - k_1}{k_2}}}$$

$$= \sqrt{\frac{k_2}{2k_2 - k_1}}$$

Is this wrong? Was an error made? Is the work term tractable but incorrect? We have assumed that $k_2$ is greater than $k_1$. The expected result, $\sqrt{k_1/k_2}$, is for $k_1 < k_2$, and $k_2 < 2k_2 - k_1$ is of this form. Note that if $\frac{1}{k_2}$, in the last line of Eq. (27), had been $\frac{1}{k_1}$ then the result would have been $\sqrt{\frac{k_1}{k_2}}$ as expected.

There is a reason for this turn of events. To see it we benefit from another calculation first. Consider the transition from $k_1$ to $k_2 < k_1$. This is a decrease in $k$ and is represented in the potential energy of the harmonic oscillator by a change in Eq. (5) (here $k_2 < k_1$ whereas in Eq. (5), $k_2 > k_1$)
(29)

$$k \to k_1 + (k_2 - k_1)\frac{t}{\tau}$$

In Eq. (23) this results in a factor of $(k_1 - k_2) > 0$ in place of its reverse, $(k_1 - k_2) < 0$. Eq. (11) is now
(30)

$$\bar{A}(s) := D\beta \left( k_1 s + (k_2 - k_1)\frac{s^2}{2\tau} \right)$$

with $(k_1 - k_2) > 0$.

In Eq. (23), the coefficient of each term in the sum is $(-1)^n$ because there is an overall factor of $-\beta$ and $k_2 > k_1$. Moreover, $W$ is quadratic in $x$. In the present case, every term is positive. One interpretation of this difference is to say that $W$ corresponds to work done *on* the oscillator and that $-W = \overline{W}$ corresponds to work, $\overline{W}$, done *by* the oscillator. With this perspective we write
(31)
$$\{\langle \exp[-\beta \overline{W}_F(\tau)]\rangle\} = \sum_{n=0}^{\infty} \left(-\frac{1}{2}(k_1 - k_2)\right)^n \frac{1}{n!} \frac{(2n)!}{n! \, 2^n} (X^2)^n$$

$$= \sum_{n=0}^{\infty} \left(-\frac{1}{2}(k_1 - k_2)\right)^n \frac{1}{n!} \frac{(2n)!}{n! \, 2^n} \left(\frac{1}{k_2}\right)^n$$

$$= \frac{1}{\sqrt{1 + \frac{k_1 - k_2}{k_2}}}$$

$$= \sqrt{\frac{k_2}{k_1}}$$

This result says that the work done *by* the oscillator on the reservoir, $\overline{W}$, is negative so that the exponential on the left-hand-side of Eq. (31) has a positive argument. Thus, the averaged exponential is greater than one. This is consistent with the final expression in Eq. (31) for $k_2 > k_1$. In summary, JE is justified for the work done *by* the oscillator but not for the work done *on* the oscillator.

What is the cause of this curious behavior? We have avoided giving away the reason for this, till now, in the hope that the reader would catch on. The harmonic oscillator *internal energy is independent of the value of $k$*. The equipartition of energy theorem states that for a harmonic oscillator *at a fixed temperature* the average kinetic energy and the average potential energy are equal. For a simple one-dimensional oscillator, the two averages are equal to $\frac{1}{2} k_B T$. For the potential energy, the average is
(32)
$$\tfrac{1}{2} k <x^2> = \tfrac{1}{2} k_B T$$

Several times in the paper to this point, this last equation has appeared in the equivalent form
(33)

$$< x^2 > = \frac{1}{\beta k}$$

This average is the result of the normalized Boltzmann distribution for the oscillator.

(34)
$$P_B(x)dx = \sqrt{\frac{\beta k}{2\pi}} \, Exp\left[-\beta \tfrac{1}{2}kx^2\right] dx$$

This means that for an isothermal process, the average of the potential energy is independent of $k$. Nevertheless, if an oscillator has a $k$ that can spontaneously change, it will change until it has reached a minimum possible value. This is because even though the internal energy remains constant the entropy does not, nor does the Helmholtz free energy. For the free energy we have

(35)
$$F = -k_B T \ln\left(\sqrt{\frac{2\pi}{\beta k}}\right)$$

and for the entropy we have

(36)
$$S = k_B \ln\left(\sqrt{\frac{2\pi}{\beta k}}\right)$$

In a spontaneous, isothermal process, an increase in $k$ would violate the second law of thermodynamics: *the (isothermal) Helmholtz free energy achieves a minimum in thermal equilibrium.* In addition, the homogeneous equation for the isothermal oscillator is

(37)
$$F = -TS$$

which means the second law for an isothermal oscillator is stated as: *the (isothermal) entropy achieves a maximum in thermal equilibrium.* This last statement is easiest to interpret physically. As the $k$ gets larger, the average excursion of $x$, given in Eq. (33), gets smaller. An oscillator with a stronger force constant keeps the position, $x$, closer to the origin than does one with a weaker force constant. This means that the effective volume available to the oscillator decreases as the force constant increases. In one dimension, the effective volume is $\sqrt{<x^2>}$. The translational part of the entropy is always of the form S $=k_B \ln(volume)$. Thus, *if* it is possible for the force constant, $k$, to spontaneously change for an oscillator in equilibrium with a thermal reservoir, then $k$ will spontaneously decrease.

These facts about the harmonic oscillator imply that a change in the spring constant, $k$, is not the same as doing work *on* the oscillator. Instead, it is a change in the entropy of the oscillator that is achieved. Even though no net work is done on the oscillator by changing the value of $k$, Eq. (31) demonstrates that the work done *by* the oscillator does satisfy JE.

## The Jarzynski Equality for the Harmonic Oscillator

The harmonic oscillator is a special case because its internal energy is independent of the force constant $k$. Thus, the work done, *on* or *by* the oscillator, expressed either by the Jarzynski work term in Eq. (7) or by Fox's work term in Eq. (8), cannot generate a change in internal energy. Consequently, we were able to verify JE only from the perspective of work done *by* the oscillator.

However, we did not obtain this restricted JE for all rates of change. Only when $\tau$ is large compared to $\tau = 7 \times 10^{-8} s$, is the asymptotic value for Eq. (27) overwhelmingly dominant. For example, any value for $\tau \in (10^{-5}s, 1s)$ or larger will imply JE. But if we explore $\tau$ for values closer to $\tau = 7 \times 10^{-8} s$, or even shorter, then many more terms in Eq. (27) become important (for an asymptotic series there is an optimal number of terms, beyond which other means of evaluation become necessary) and the results are far from compatible with JE. Realistic experiments are very likely to be performed only in the larger $\tau$ regime, apparently justifying the claim that experiments have confirmed JE for arbitrary rates. For the rates where there is a difference, $\tau$ is much smaller and experiments much harder. A systematic review of all experiments performed so far is called for by this analysis.

Jarzynski [10] has shown me a beautiful derivation of JE starting from the Smoluchowski equation. See appendix B. In his derivation, it appears to be the case that JE is true for all rates. As above, where a short enough $\tau$ will invalidate this claim, the validity of the Smoluchowski equation also has a limit. At short enough times the separation of momentum space and coordinate space relaxation times breaks down and *the Smoluchowski equation is no longer valid*. It is the Markov approximation (overdamped diffusion limit) to a more detailed dynamical description [6, 7]. Thus, the conclusion that the form of the solution found in appendix B works for any rate whatsoever is not true when the regime of validity of the Smoluchowski equation *itself* is exceeded.

Hummer and Szabo [11] have argued that JE follows easily from the Feynman-Kac formula [12]. A superficial inspection makes this claim seductively tantalizing. In my correspondence with Jarzynski, I argued that the application of F-K to JE overlooked a difference in a boundary condition for one of the key equations. Hummer and Szabo (as well as Jarzynski) used a distribution function boundary condition that is proportional to the Boltzmann distribution (un-normalized) whereas in F-K the corresponding boundary condition is a Dirac delta function [12]. This difference clearly invalidates the application of F-K to the derivation of JE. In our correspondence, Jarzynski made it clear that he has assiduously avoided invoking F-K in his own derivations.

As reinforcement of the result in Eq. (31), look at the process in which a *decrease* in $k$ is only partially completed. That is, we stop at time $t < \tau$ in which the inequality is strict. Returning to Eq. (31), the restriction to $t < \tau$ for the work done *by* the oscillator is expressed by

(38)
$$\{\langle \exp[-\beta \overline{W}_F(t)] \rangle\} = \sum_{n=0}^{\infty} \left(-\frac{1}{2}(k_1 - k_2)\frac{t}{\tau}\right)^n \frac{1}{n!} \frac{(2n)!}{n! \, 2^n} (X^2)^n$$

in which $(k_1 - k_2) > 0$ and $x^2(t)$ is given by

(39)
$$\{\langle x^2(t) \rangle\}$$

$$= \exp[-2A(t)] \left(\frac{1}{\beta k_1} + \int_0^t ds \, \exp[2A(s)] 2D\right)$$

$$= \exp\left[-2D\beta\tau\left(k_1 \frac{t}{\tau} + (k_2 - k_1)\frac{t^2}{2\tau^2}\right)\right] \left(\frac{1}{\beta k_1}\right.$$

$$+ \frac{exp\left[-\frac{D\beta\tau k_1^2}{(k_2 - k_1)}\right]\sqrt{\pi D\beta\tau}\left(Erf\left[\frac{\sqrt{D\beta\tau}k_1}{\sqrt{(k_1 - k_2)}}\right] - Erf\left[\frac{\sqrt{D\beta\tau}\left(k_2\frac{t}{\tau} + k_1\left(1 - \frac{t}{\tau}\right)\right)}{\sqrt{(k_1 - k_2)}}\right]\right)}{\beta\sqrt{(k_1 - k_2)}}\right)$$

We are calling this $\{\langle x^2(t)\rangle\}$ rather than $X^2(t)$ because we haven't yet canceled out the numerator and denominator $\beta's$. Invoke Eq. (26) to approximate the asymptotic values for these quantities. Note that for $(k_1 - k_2) > 0$, the arguments of the Erf functions are real and rapidly reach their asymptotic boundaries of $\pm 1$. This yields

(40)
$$\{\langle x^2(t)\rangle\}$$

$$= exp\left[-2D\beta\tau\left(k_1\frac{t}{\tau} + (k_2 - k_1)\frac{t^2}{2\tau^2}\right)\right]\left(\frac{1}{\beta k_1}\right.$$

$$+ \frac{exp\left[-\frac{D\beta\tau k_1^2}{(k_2 - k_1)}\right]\sqrt{\pi D\beta\tau}\left(Erf\left[\frac{\sqrt{D\beta\tau}k_1}{\sqrt{(k_1 - k_2)}}\right] - Erf\left[\frac{\sqrt{D\beta\tau}\left(k_2\frac{t}{\tau} + k_1\left(1 - \frac{t}{\tau}\right)\right)}{\sqrt{(k_1 - k_2)}}\right]\right)}{\beta\sqrt{(k_1 - k_2)}}\right)$$

$$\approx Exp\left[-2D\beta\tau\left(k_1\frac{t}{\tau}-(k_1-k_2)\frac{t^2}{2\tau^2}\right)\right]\left(\frac{1}{\beta k_1}\right.$$

$$+\frac{exp\left[\frac{D\beta\tau k_1^2}{(k_1-k_2)}\right]\sqrt{\pi D\beta\tau}}{\beta\sqrt{(k_1-k_2)}}\left(\left(-\frac{1}{\sqrt{\pi}}\right)\left(\left(\frac{\sqrt{(k_1-k_2)}}{\sqrt{D\beta\tau}k_1}\right)Exp\left[-\frac{D\beta\tau k_1^2}{(k_1-k_2)}\right]\right.\right.$$

$$\left.\left.-\left(\frac{\sqrt{(k_1-k_2)}}{\sqrt{D\beta\tau}\left(k_2\frac{t}{\tau}+k_1\left(1-\frac{t}{\tau}\right)\right)}\right)Exp\left[-\frac{D\beta\tau\left(k_2\frac{t}{\tau}+k_1\left(1-\frac{t}{\tau}\right)\right)^2}{(k_1-k_2)}\right]\right)\right)\right)$$

$$=\frac{1}{\beta\left(k_2\frac{t}{\tau}+k_1\left(1-\frac{t}{\tau}\right)\right)}$$

As $t$ goes from 0 to $\tau$, this expression linearly interpolates an expression for $k$ going from $k_1$ to $k_2$. Eq. (38) takes the form

(41)
$$\{\langle\exp[-\beta\overline{W}_F(t)]\rangle\}=\sum_{n=0}^{\infty}\left(-\frac{1}{2}(k_1-k_2)\frac{t}{\tau}\right)^n\frac{1}{n!}\frac{(2n)!}{n!\,2^n}\left(\frac{1}{\left(k_2\frac{t}{\tau}+k_1\left(1-\frac{t}{\tau}\right)\right)}\right)^n$$

$$=\frac{1}{\sqrt{1+\frac{(k_1-k_2)\frac{t}{\tau}}{\left(k_2\frac{t}{\tau}+k_1\left(1-\frac{t}{\tau}\right)\right)}}}$$

$$=\frac{1}{\sqrt{\frac{\left(k_2\frac{t}{\tau}+k_1\left(1-\frac{t}{\tau}\right)\right)+(k_1-k_2)\frac{t}{\tau}}{\left(k_2\frac{t}{\tau}+k_1\left(1-\frac{t}{\tau}\right)\right)}}}$$

$$= \frac{1}{\sqrt{\frac{k_1}{\left(k_2 \frac{t}{\tau} + k_1\left(1 - \frac{t}{\tau}\right)\right)}}}$$

$$= \sqrt{\frac{k_2 \frac{t}{\tau} + k_1\left(1 - \frac{t}{\tau}\right)}{k_1}}$$

This is JE for $k$ starting at $k_1$ and ending at time $t < \tau$, for the case $(k_1 - k_2) > 0$. That means that the numerator in the last line is smaller than the denominator as is expected from the form of the left-hand side of Eq. (41) if $\overline{W}_F$ is positive.

## Conclusions and Acknowledgments

My friend and colleague at The Rockefeller University, Eddie Cohen, exposed me to JE just after the turn of the century. He was then just about my age now (he died in 2017 at the age of 94). Eddie had many objections and he carried out a vigorous campaign probing JE in a series of publications [4]. Jarzynski patiently defended himself and parried each attack [5].

My experience in pursuing this matter has been very positive. Jarzynski has exhibited the patience of Job. The identity in Eq. (15) compels one to the view that we are on the right track and that we are not done yet. For the oscillator spring constant, it appears to be the entropy rather than the work that is behind JE. JE is not true for arbitrarily fast rates. The oscillator still may have more to teach us. When true, why is JE true?

# Appendix A

Let X and Y be independent gaussian variables with zero means and with standard deviations $\sigma_X$ and $\sigma_Y$ respectively. These properties are expressed with the notations

$$<X> = 0$$

$$<Y> = 0$$

$$<XY> = 0$$

$$<X^2> = \sigma_X^2$$

$$<Y^2> = \sigma_Y^2$$

For a gaussian variable with zero mean, an equivalent characterization as gaussian is given by the formula

$$<X^{2n}> = \frac{(2n)!}{2^n \, n!}(\sigma_X^2)^n$$

Lemma: With X and Y characterized as above, $X + Y$ is also gaussian, or equivalently,

$$<(X+Y)^{2n}> = \frac{(2n)!}{2^n \, n!}(\sigma_X^2 + \sigma_Y^2)^n$$

Proof:

$$<(X+Y)^{2n}> = <\sum_{m=0}^{n}\frac{(2n)!}{(2m)!\,(2n-2m)!}\left(X^{2(n-m)}Y^{2m}\right)>$$

$$= \sum_{m=0}^{n}\frac{(2n)!}{(2m)!\,(2n-2m)!}<X^{2(n-m)}><Y^{2m}>$$

$$= \sum_{m=0}^{n}\frac{(2n)!}{(2m)!\,(2n-2m)!}\frac{(2n-2m)!}{2^{n-m}\,(n-m)!}(<X^2>)^{n-m}\frac{(2m)!}{2^m\,m!}(<Y^2>)^m$$

$$= \sum_{m=0}^{n}\frac{(2n)!}{2^n\,(n-m)!\,m!}(\sigma_X^2)^{n-m}(\sigma_Y^2)^m$$

$$= \frac{(2n)!}{2^n\,n!}\sum_{m=0}^{n}\frac{(n)!}{(n-m)!\,m!}(\sigma_X^2)^{n-m}(\sigma_Y^2)^m$$

$$= \frac{(2n)!}{2^n\,n!}(\sigma_X^2 + \sigma_Y^2)^n$$

QED

# Appendix B

The exchange of ideas I had with Christopher Jarzynski over several weeks in 2021 was unexpected when I initially sent him a preprint. Instead of the swift dismissal I sometimes experienced with other scientists in other fields, I got back a carefully reasoned pedagogically structured treatment of the problem put into the notation I used in my preprint rather than the usual notation used by Jarzynski in his papers. The argument in favor of JE given in this appendix is his argument as sent to me during our exchange. I present it here partly because of its convincing structure but mostly because it permits a direct rendering of the question: is JE really independent of the rate of transition? Several other issues remain to be given greater clarification. Chief among them is a deeper understanding of how well JE comes to being justified even though the harmonic oscillator case is a special one in that there is no change in internal energy as $k$ changes, only changes in entropy. The agreement with JE is better when the change in spring constant is in the direction of changes in the entropy consistent with the second law. Otherwise, changes forced in the direction contrary to the second law are problematic.

## Jarzynski's proof of JE for the Harmonic Oscillator

These comments refer to emails exchanged on August 3, 2021, and August 5, 2021. The active, first-person voice may be taken to be Chris's, although in fact Ron has transcribed it here. Ron's editorial comments are found inside the {...}'s.

Consider a particle evolving under the Langevin equation
(B1)
$$\frac{dx}{dt} = -\beta D U'(x,t) + g(t)$$

where $g(t)$ is the fluctuating force described by Eq. 3 of your draft, and $U' = \partial_x U$. The corresponding Fokker-Planck equation for the probability density $f(x,t)$ is
(B2)
$$\frac{\partial f}{\partial t} = \beta D \frac{\partial}{\partial x}(U'f) + D \frac{\partial^2 f}{\partial x^2}$$

which is equivalent to your Eq. 1. { $\frac{\partial R}{\partial t} = D \frac{\partial}{\partial x}\left(\frac{\partial}{\partial x} + \beta U'\right) R$ }

Now let $w(\tau)$ denote the work performed on the system from time 0 to time $\tau$, hence $w(0) = 0$ and $w(\tau)$ is equal to your $W$. {Chris is referring to Eq. (7)

above which is also his $W_J$ and is distinct from Ron's $W_F$ in Eq. (8).} The evolving quantities $x(t)$ and $w(t)$ obey the coupled equations of motion

(B3a)
$$\frac{dx}{dt} = -\beta D U'(x,t) + g(t)$$

(B3b)
$$\frac{dw}{dt} = \dot{U}(x,t)$$

where $\dot{U} = \partial_t U$. {This equation holds for both definitions in Eqs. (7, 8)} (Eq. B3a is just Eq. 1). Letting $h(x,w,t)$ denote the joint probability density for $x$ and $w$, we have {!}

(B4)
$$\frac{\partial h}{\partial t} = \beta D \frac{\partial}{\partial x}(U'h) + D\frac{\partial^2 h}{\partial x^2} - \dot{U}\frac{\partial h}{\partial w}$$

The first two terms on the right describe the drift and the diffusion in the x-direction, as in Eq. B2 above, and the third term is a continuity term that arises from Eq. B3b. Eqs. B3 and B4 extend Eqs. B1 and B2 from $x$-space to $(x,w)$-space.

Now define {!!}

(B5)
$$g(x,t) \equiv \int dw \, e^{-\beta w} h(x,w,t)$$

so that

(B6)
$$<e^{-\beta W}> = \int dx \, g(x,\tau)$$

The equation of motion for $g(x,t)$ follows from Eqs. B4 and B5, using integration by parts: {!!!}

(B7)
$$\frac{\partial g}{\partial t} = \int dw \, e^{-\beta w} \frac{\partial h}{\partial t} = \beta D \frac{\partial}{\partial x}(U'g) + D\frac{\partial^2 g}{\partial x^2} - \beta \dot{U} g$$

Since $w(0) = 0$, the initial conditions for $g(x,t)$, are the same as those for $f(x,t)$:

(B8)
$$g(x,0) = f(x,0) = \frac{1}{Z_0} e^{-\beta U(x,0)}$$

where $Z_0 = \int dx\, e^{-\beta U(x,0)}$ is the partition function associated with $U(x,0)$.

From the initial conditions given by Eq. B8 it is easy to verify, by direct substitution, that the (unique) solution of Eq. B7 is
(B9) {!!!!}
$$g(x,t) = \frac{1}{Z_0} e^{-\beta U(x,t)}$$

Eq. B6 immediately gives $\{t \to \tau\}$
(B10)
$$< e^{-\beta W} > = \frac{Z_\tau}{Z_0} = e^{-\beta \Delta F}$$

QED